\documentclass[12pt,a4paper]{article}
\usepackage{amsmath}
\usepackage{graphicx}
\begin{document}
\title{
Ward identities for extended objects
}

\author{
O. Narikiyo
\footnote{
Department of Physics, 
Kyushu University, 
Fukuoka 812-8581, 
Japan}
}

\date{
(Dec. 28, 2012)
}

\maketitle
\begin{abstract}
Ward identities for extended objects are discussed. 
In the limit of dc transport 
it is rigorously proved that 
charge-density and spin-density fluctuations do not couple 
to electromagnetic field.
\end{abstract}

\vskip 30pt 

\section{Introduction}

In most cases of condensed matter physics 
Ward identities have been discussed 
for single-particle vertecies~\cite{Sch,Ono}. 
Recently I have reported the discussions of Ward identities 
for particle-particle or particle-hole pairs~\cite{W1,W2,W3}. 

In this note I will summarize the essential point for extended pairs. 
Following description is based on refs.~\cite{W1,W2,W3}. 

\section{Ward identity for electric current vertex}

The central quantity in the discussions of Ward identity 
for electric current vertex 
is the equal-time commutation relation 
between the electric charge 
\begin{equation}
j_{\vec k}^e = e 
\sum_{\vec p'} 
\Bigl( a_{{\vec p'}-{\vec k}}^\dag a_{\vec p'} 
     + b_{{\vec p'}-{\vec k}}^\dag b_{\vec p'} \Bigr), 
\end{equation}
and the object which comes in or go out of the vertex. 

If the object is an electron, 
then the equal-time commutation relation is estimated as 
\begin{equation}
[ j_{\vec k}^e, a_{\vec p}^\dag ] = e a_{{\vec p}-{\vec k}}^\dag, 
\end{equation}
where $a_{\vec p}^\dag$ is the creation operator of an electron. 
If the object is a local Cooper pair, 
\begin{equation}
[ j_{\vec k}^e, P_{\vec q}^\dag ] = 2 e P_{{\vec q}-{\vec k}}^\dag, 
\end{equation}
where $P_{\vec q}^\dag$ is the creation operator of a pair. 
In the right-hand side of these equations 
the charge carried by the object appears; 
$e$ in the case of the electron and $2e$ in the case of the Cooper pair. 

Next we consider the case of extended Cooper pairs. 
An extended Cooper pair is represented as 
\begin{equation}
\Psi({\vec R}) = 
\int d {\vec r} \chi_l({\vec r}) 
\psi_\downarrow({\vec r_1}) \psi_\uparrow({\vec r_2}), 
\end{equation}
where 
\begin{equation}
\chi_l({\vec r}) = 
\sum_{\vec p} e^{i{\vec p}\cdot{\vec r}} \chi_l({\vec p}),\ \ \ 
\psi_\downarrow({\vec r_1}) = 
\sum_{\vec p_1} e^{i{\vec p_1}\cdot{\vec r_1}} b_{\vec p_1},\ \ \ 
\psi_\uparrow({\vec r_2}) = 
\sum_{\vec p_2} e^{i{\vec p_2}\cdot{\vec r_2}} a_{\vec p_2}. 
\end{equation}
Here ${\vec R}$ is the center-of-mass coordinate of the pair 
and ${\vec r}$ is the relative coordinate. 
Namely, ${\vec r_1}={\vec R}+{\vec r}/2$ and ${\vec r_2}={\vec R}-{\vec r}/2$. 
Performing the integral in terms of the relative coordinate ${\vec r}$ 
we obtain\footnote{
Eq.~(18) in ref.~\cite{W2} should be replaced by (6) in this note. 
} 
\begin{equation}
\Psi({\vec R}) = 
\sum_{\vec q} e^{i{\vec q}\cdot{\vec R}} P_{\vec q}, 
\end{equation}
with 
\begin{equation}
P_{\vec q} = 
\sum_{\vec p} \chi_l({\vec p}) 
b_{{{\vec q} \over 2}-{\vec p}} a_{{{\vec q} \over 2}+{\vec p}}, \ \ \ \ \ 
P_{\vec q}^\dag = 
\sum_{\vec p} \chi_l({\vec p}) 
a_{{{\vec q} \over 2}+{\vec p}}^\dag b_{{{\vec q} \over 2}-{\vec p}}^\dag. 
\end{equation}

The equal time commutation relation is calculated as 
\begin{align}
[ j_{\vec k}^e, P_{\vec q}^\dag ] 
= & e \sum_{\vec p'} \sum_{\vec p} \chi_l({\vec p}) 
\Bigl( 
a_{{\vec p'}-{\vec k}}^\dag \Big\{ 
a_{\vec p'} a_{{{\vec q} \over 2}+{\vec p}}^\dag + 
a_{{{\vec q} \over 2}+{\vec p}}^\dag a_{\vec p'} 
\Big\} b_{{{\vec q} \over 2}-{\vec p}}^\dag \nonumber
\\ 
& \ \ \ \ \ \ \ \ \ \ \ \ \ \ \ \ \ \ - 
b_{{\vec p'}-{\vec k}}^\dag \Big\{ 
b_{\vec p'} b_{{{\vec q} \over 2}-{\vec p}}^\dag + 
b_{{{\vec q} \over 2}-{\vec p}}^\dag b_{\vec p'} 
\Big\} a_{{{\vec q} \over 2}+{\vec p}}^\dag \nonumber
\\ 
= & e \sum_{\vec p} \chi_l({\vec p}) 
\Bigl( 
a_{{{\vec q} \over 2}+{\vec p}-{\vec k}}^\dag 
b_{{{\vec q} \over 2}-{\vec p}}^\dag 
+ 
a_{{{\vec q} \over 2}+{\vec p}}^\dag 
b_{{{\vec q} \over 2}-{\vec p}-{\vec k}}^\dag 
\Bigr) \nonumber
\\ 
= & e \sum_{\vec p} \chi_l({\vec p}) 
\Bigl( 
a_{{{\vec q}-{\vec k} \over 2}+({\vec p}-{{\vec k} \over 2})}^\dag 
b_{{{\vec q}-{\vec k} \over 2}-({\vec p}-{{\vec k} \over 2})}^\dag 
+ 
a_{{{\vec q}-{\vec k} \over 2}+({\vec p}+{{\vec k} \over 2})}^\dag 
b_{{{\vec q}-{\vec k} \over 2}-({\vec p}+{{\vec k} \over 2})}^\dag 
\Bigr). 
\end{align}

By shifting the variable of the summation we obtain 
\begin{equation}
[ j_{\vec k}^e, P_{\vec q}^\dag ] \rightarrow 2 e 
\sum_{\vec p'} \chi_l({\vec p'}) 
a_{{{\vec q}-{\vec k} \over 2}+{\vec p'}}^\dag 
b_{{{\vec q}-{\vec k} \over 2}-{\vec p'}}^\dag 
= 2 e P_{{\vec q}-{\vec k}}^\dag, 
\end{equation}
in the limit of vanishing external momentum ${\vec k}\rightarrow 0$, 
where the shift in the argument of the form factor 
$\chi_l({\vec p'}\pm{{\vec k}\over 2})$ is negligible. 
Thus even in the case of extended Cooper pair 
the commutation relation in the limit of vanishing external momentum 
picks up the integrated charge of the object 
so that we obtain the same Ward identity as in the case of local pair. 

Next we consider the case of extended particle-hole pairs. 
An extended particle-hole pair is represented as 
\begin{equation}
A^\dag({\vec R}) = 
\int d {\vec r} \chi({\vec r}) 
\psi_\uparrow^\dag({\vec r_1}) \psi_\downarrow({\vec r_2}), 
\end{equation}
where 
\begin{equation}
\chi({\vec r}) = 
\sum_{\vec p} e^{i{\vec p}\cdot{\vec r}} \chi({\vec p}),\ \ \ 
\psi_\uparrow^\dag({\vec r_1}) = 
\sum_{\vec p_1} e^{-i{\vec p_1}\cdot{\vec r_1}} a_{\vec p_1}^\dag,\ \ \ 
\psi_\downarrow({\vec r_2}) = 
\sum_{\vec p_2} e^{i{\vec p_2}\cdot{\vec r_2}} b_{\vec p_2}. 
\end{equation}
Performing the integral in terms of the relative coordinate ${\vec r}$ 
we obtain 
\begin{equation}
A^\dag({\vec R}) = 
\sum_{\vec q} e^{i{\vec q}\cdot{\vec R}} A_{\vec q}^\dag, 
\end{equation}
with 
\begin{equation}
A_{\vec q}^\dag = 
\sum_{\vec p} \chi({\vec p}) 
a_{{\vec p}-{{\vec q} \over 2}}^\dag 
b_{{\vec p}+{{\vec q} \over 2}}. 
\end{equation}

The equal time commutation relation is calculated as 
\begin{align}
[ j_{\vec k}^e, A_{\vec q}^\dag ] 
= & e \sum_{\vec p'} \sum_{\vec p} \chi({\vec p}) 
\Bigl( 
a_{{\vec p'}-{\vec k}}^\dag \Big\{ 
a_{\vec p'} a_{{\vec p}-{{\vec q} \over 2}}^\dag + 
a_{{\vec p}-{{\vec q} \over 2}}^\dag a_{\vec p'} 
\Big\} b_{{\vec p}+{{\vec q} \over 2}} \nonumber
\\ 
& \ \ \ \ \ \ \ \ \ \ \ \ \ \ \ \ \ \ - 
a_{{\vec p}-{{\vec q} \over 2}}^\dag \Big\{ 
b_{{\vec p'}-{\vec k}}^\dag b_{{\vec p}+{{\vec q} \over 2}} + 
b_{{\vec p}+{{\vec q} \over 2}} b_{{\vec p'}-{\vec k}}^\dag 
\Big\} b_{{\vec p'}} \nonumber
\\ 
= & e \sum_{\vec p} \chi({\vec p}) 
\Bigl( 
a_{{\vec p}-{{\vec q} \over 2}-{\vec k}}^\dag 
b_{{\vec p}+{{\vec q} \over 2}} 
- 
a_{{\vec p}-{{\vec q} \over 2}}^\dag 
b_{{\vec p}+{{\vec q} \over 2}+{\vec k}} 
\Bigr) \nonumber
\\ 
= & e \sum_{\vec p} \chi({\vec p}) 
\Bigl( 
a_{({\vec p}-{{\vec k} \over 2})-{{\vec q}+{\vec k} \over 2}}^\dag 
b_{({\vec p}-{{\vec k} \over 2})+{{\vec q}+{\vec k} \over 2}} 
- 
a_{({\vec p}+{{\vec k} \over 2})-{{\vec q}+{\vec k} \over 2}}^\dag 
b_{({\vec p}+{{\vec k} \over 2})+{{\vec q}+{\vec k} \over 2}} 
\Bigr). 
\end{align}

By the same procedure as in the case of the Cooper pair 
we obtain
\begin{equation}
[ j_{\vec k}^e, A_{\vec q}^\dag ] \rightarrow e 
\sum_{\vec p'} \chi({\vec p'}) \Bigl( 
a_{{\vec p'}-{{\vec q}+{\vec k} \over 2}}^\dag 
b_{{\vec p'}+{{\vec q}+{\vec k} \over 2}} 
- 
a_{{\vec p'}-{{\vec q}+{\vec k} \over 2}}^\dag 
b_{{\vec p'}+{{\vec q}+{\vec k} \over 2}} \Bigr) 
= 0. 
\end{equation}
Since the integrated charge of the particle-hole pair vanishes, 
the commutation relation vanishes. 
The current vertex for dc conductivity 
is obtained from the Ward identity 
in the limit of vanishing external momentum $ k \rightarrow 0 $. 
Thus the Ward identity derived from this commutation relation tells us 
that particle-hole pairs have no contribution to dc conductivity. 
Such a conclusion is natural, 
since particle-hole pairs are charge-neutral and carry no charge. 

\section{Ward identity for heat current vertex}

The central quantity in the discussions of Ward identity 
for heat current vertex 
in the limit of vanishing external momentum $ k \rightarrow 0 $ 
is the equal-time commutation relation 
between the Hamiltonian and the object 
which comes in or go out of the vertex. 
As the discussion in the previous section, 
we obtain the same Ward identity as in the case of local pair 
in this limit. 

\section{Conclusion}

In the discussion of the Ward identity 
an equal-time commutation relation plays the central role. 

In the case of the electric current vertex 
it picks up the integrated charge of the object 
in the limit of vanishing external momentum. 
In this limit 
the wavelength of the electromagnetic field exceeds 
the size of the object 
so that the object can be treated as a point with its integrated charge 
in the discussion of the electromagnetic response. 
Thus it is concluded that charge-neutral pairs do not couple 
to electromagnetic field. 
Namely, 
charge- and spin-density fluctuations do not carry charge. 
On the other hand, 
Cooper pairs carrying charge $2e$ 
couple to electromagnetic field. 

In the case of the heat current vertex 
it picks up the energy of the object. 

\newpage


\end{document}